\documentclass[twocolumn,aps,prl,floatfix,showpacs]{revtex4}
\usepackage{graphicx}
\begin{document}

\title{Dispersion anomalies induced by the low-energy plasmon in the 
cuprates}

\author{R.S. Markiewicz and A. Bansil}
\affiliation{
Physics Department, Northeastern University, Boston MA 02115}
\date{\today}
\begin{abstract}

We discuss the characteristic effects of the electron plasmon interaction 
resulting from the $\sim$ 1 eV plasmon, which is a universal feature in 
the cuprates. Using the framework of a one-band tight binding model, we 
identify signatures of this low energy plasmon in the electronic structure 
of metallic overdoped Bi2212 as well as half-filled insulating SCOC. The 
electron-plasmon interaction is found to yield renormalizations near the 
Fermi energy in reasonable accord with experimental observations, and to 
produce dispersion anomalies at higher energies.

\end{abstract}
\maketitle

Quasiparticle dispersions near the Fermi energy in the cuprates are 
renormalized (reduced) by a factor of $Z\sim 0.3-0.6$\cite{Arun3} in 
comparison to the band theory predictions based on the conventional LDA 
picture, interrupted by the appearance of low energy `kinks' or dispersion 
anomalies in the 50-70 meV energy range, which arise from the coupling of 
the electronic system with phonons\cite{pkink} and/or magnetic 
modes\cite{mkink}. Interestingly, the `spaghetti' of various hybridized 
and unhybridized Cu and O bands starting around 1 eV in the cuprates seems 
to remain more or less unrenormalized.\cite{Ale1} Recalling that 
optical\cite{Boz} and electron energy loss spectroscopy (EELS) 
experiments\cite{HedLee,WZD} have shown the presence of dispersive 
plasmons in the cuprates lying at $\sim$1 eV, we explore the effects of 
such a low energy plasmon in this Letter. The plasmon is found to induce 
renormalizations in the low energy regime in both the metallic and the 
insulating cuprates, which are in substantial accord with the 
corresponding experimental results. We emphasize that plasmons in 3D 
materials typically lie at $\sim$10 eV and therefore have little influence 
on the electronic states in the 0-1 eV range.

We approach the problem within the framework of a one-band tight-binding 
model and consider the limiting cases of a metal with the example of 
overdoped Bi$_2$Sr$_2$CaCu$_2$O$_{8}$ (Bi2212)\cite{Ale,Non} and the 
insulator with the example of half-filled Sr$_2$CuO$_2$Cl$_2$ 
(SCOC)\cite{RonK}. The dielectric functions computed via the random phase 
approximation (RPA) yield loss functions and the associated plasmon 
dispersions in both Bi2212 and SCOC, which are in reasonable accord with 
the corresponding EELS data. The generic effects of the electron-plasmon 
interaction on the electronic structure and how the low energy plasmon in 
the metal and the insulator induces characteristic dispersion anomalies 
and renormalizations of the spectrum are then delineated via a first order 
calculation of the electronic self energy.

We discuss first the metallic case of overdoped Bi2212. 
Here the magnetic gap has collapsed fully and we take the bare electronic 
dispersion to be given by a one-band tight-binding model fitted to the 
LDA-based band structure of Bi2212 in the vicinity of the Fermi energy 
($E_F$) as follows\cite{Arun3}:
\begin{eqnarray}
     \epsilon_{\bf k}=-2t[c_x(a)+c_y(a)]-4t'c_x(a)c_y(a) 
     -2t''[c_{x}(2a)
\nonumber \\
+c_{y}(2a)] 
     -4t'''[c_{x}(2a)c_y(a)+c_{y}(2a)c_x(a)] \> , \label{eq:A2} 
\end{eqnarray} 
where $c_\alpha (na)=cos(nk_\alpha a)$ for $\alpha=x$ or 
$y$, $n$ is an integer, and $a$ is the in-plane lattice constant. Hopping parameters 
appropriate for Bi2212 (neglecting bilayer splitting)\cite{Arun3} are: $t$=360 meV, 
$t'=-100$ meV, $t''$=35 meV, $t'''=10 meV$. We model the dynamic dielectric function, 
$\epsilon(q,\omega)$, in the form\cite{HedLee,static} 
\begin{equation} 
\epsilon(q,\omega) =\epsilon_0\Bigl[1-{\omega^2_p(q)\over \omega 
(\omega-i\gamma(q) )}\Bigr], 
\label{eq:13} 
\end{equation} 
where $\epsilon_0=6$ and the dispersion and damping of the plasmon in 
Bi2212 obtained from EELS data are given by\cite{HedLee}: 
$\omega_p=1.1eV[1+4S_{xy}]$, $\gamma =0.7eV[1+16S_{xy}]$, 
$S_{xy}=sin^2(q_xa/2)+sin^2(q_ya/2)$. Notably, due to the 
large broadening, the experimental plasmon peak lies at a lower energy 
than the bare value $\omega_p$.
Using the bare dispersion of Eq.~1 and the dielectric function of Eq.~2, 
it is straightforward\cite{Mah} to obtain the first order correction to 
the electronic self energy $\Sigma$ due to the screened Coulomb 
interaction\cite{Footnote}.  The formulas can also be derived from those for the 
insulator (Eqs.~5-11 below) by letting the magnetic gap 
$\Delta_m\rightarrow 0$.

\begin{figure}  
            \resizebox{8.5cm}{!}{\includegraphics{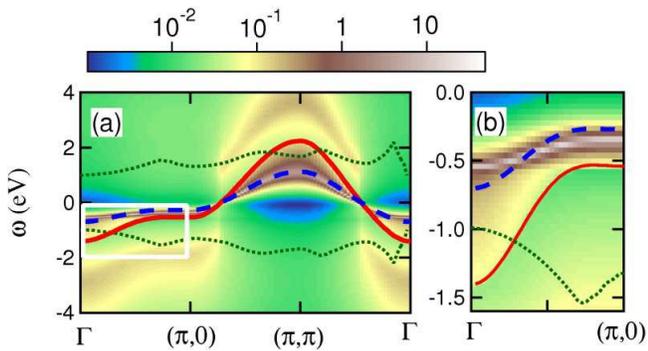}}
\vskip0.5cm
\caption{
(Color online) (a) Dressed spectral density function in overdoped 
(metallic) Bi2212.  Bare dispersion (red line) is compared with dispersion 
renormalized with $Z=0.5$ (blue dashed line). Dotted lines indicate 
plasmon dispersion.  (b) Blowup of region enclosed by white box in 
(a). Logarithmic color scale is used to reveal weaker features.
}
\label{fig:3}
\end{figure}

Figure~1 displays the spectral density in the overdoped metallic case 
where effects of the electron-plasmon interaction are included. We 
consider the low energy region along $\Gamma\rightarrow (\pi ,0)$ shown in 
(b) first. A region of high spectral intensity depicted by the thin white 
trace in (b) can be seen clearly (note logarithmic scale), indicating that 
the LDA band continues to persist in the presence of the plasmon, albeit 
with some renormalization. An average factor of $Z=0.5$ is seen to more or 
less describe the renormalization of the LDA band as seen by comparing the 
dashed blue line with the whitish trace of high intensity. The $Z=0.5$ 
value so determined is in substantial accord with the corresponding 
experimental value of $Z=0.28$,\cite{Arun3} indicating that the plasmon 
constitutes a substantial part of the bosonic dressing of the electron. 
Moreover, the spectral weight spreads out from the renormalized band 
towards the bare band (red line), giving the resulting yellowish trace the 
visual appearance of a 'waterfall' terminating around the $\Gamma$ point. 
These results bear some resemblance to 'waterfall'-like anomalies reported 
recently in the ARPES spectra from the cuprates over the 0.4-0.8 eV 
energies\cite{Ale,Non}. However, the waterfall-like features induced by 
the plasmon and those reported in ARPES differ significantly insofar as 
their location in energy and momentum is concerned, and therefore, on the 
face of it, the plasmon would not seem to provide an explanation of the 
recent ARPES results. We return to comment further on this point below.

We now turn to Fig. 1(a) which exposes effects of the plasmon over a wider 
energy scale. Here we see that the plasmon induces 'plasmaron' 
features\cite{HedLund} -- a new 'shadow' band lying 1.5 eV below the bare 
dispersion via the plasmon emission process along the filled portion of 
the LDA bands along the $\Gamma$ to $(\pi,0)$ and the $\Gamma$ to 
$(\pi,\pi)$ lines. A similar shadow band associated with plasmon 
absorption appears 1.5~eV above the unfilled part of the LDA band along 
the $(\pi,0)$ to $(\pi,\pi)$ and $\Gamma$ to $(\pi,\pi)$ lines. At low 
energies, the plasmon dresses the electrons, making them heavier and 
renormalizing their dispersion. In contrast, at high energies, the plasmon 
becomes progressively less effective in dressing the electrons which move 
with ever increasing speeds, even though the electronic states are still 
broadened by plasmon emission effects.

\begin{figure} 
            \resizebox{8.5cm}{!}{\includegraphics{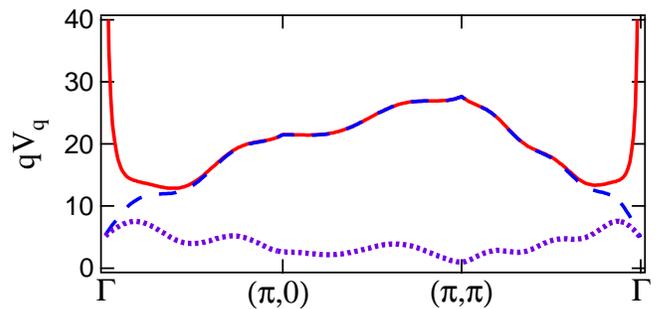}}
\vskip0.5cm
\caption{
(color online) Different calculations of the Coulomb potential (see text) 
in the form of $qV_{\bf q}$ are compared. Full 3D layered lattice (solid 
line); A single 2D layer (dashed); and 2D layer without the on-site $U$ 
term (dotted).
}
\label{fig:1}
\end{figure}

We now demonstrate that the experimentally observed plasmon dispersion 
in Bi2212 is in fact described reasonably well by the band 
structure of Eq. 1, consistent with the results of Ref.~\cite{SharK}. For 
this purpose, we directly evaluate the dielectric function from the bare 
charge susceptibility $\chi_0$ and the Coulomb interaction $V_{\bf q}$ via
\begin{equation}
\epsilon (\bf q,\omega )=\epsilon_0[1+V_{\bf q}\chi_0(\bf q,\omega )].
\label{eq:4}
\end{equation}
$\chi_0$ is calculated straightforwardly within the RPA approximation 
where the band structure of Eq. 1 is renormalized by a factor of Z=0.28 to 
account for the renormalization observed in ARPES experiments on Bi2212. 
The treatment of the long-range Coulomb interaction $V(r)$ for the 
correlated electronic system in a layered structure requires some care and 
we model $V(r)$ as an on-site Hubbard $U=$2 eV term\cite{foot0} plus a $1/r$ 
contribution screened by the background dielectric constant $\epsilon_0$ at 
other lattice sites.  $V_q$ is then obtained via a lattice Fourier transform 
by summing the contributions of all in-plane Cu terms for $r\le R\equiv 4a$ 
($a$ being the in-plane lattice constant) and approximating the remaining 
in-plane as well as interplane contributions for $r>R$ by a continuum in order 
to recover the correct $q\rightarrow 0$ limit of $V_q$. Fig.~\ref{fig:1} 
shows $qV_{\bf q}$ along high symmetry lines in the Brillouin 
zone (BZ). $V_{\bf q}$ is seen to display a crossover from 3D behavior 
$\sim 1/q^2$ to 2D behavior $\sim 1/q$ for $qc>1$, where $c$ is the 
distance between successive CuO$_2$ planes.

\begin{figure}  
            \resizebox{8.5cm}{!}{\includegraphics{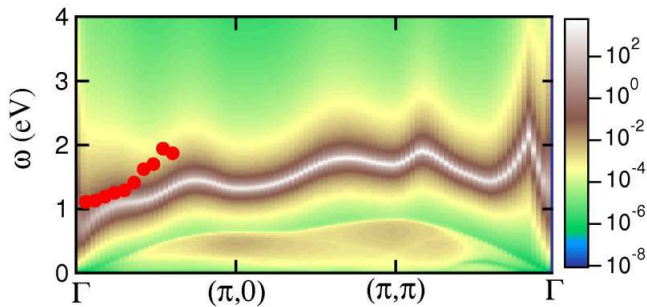}}
\vskip0.5cm
\caption{
(Color online) Loss function Im $[-1/\epsilon]$ as a function of $\omega$ 
and ${\bf q}$ in Bi2212 plotted on a logarithmic color scale. Red dots  
give the experimental plasmon peak positions \protect\cite{HedLee}.
}
\label{fig:2}
\end{figure}

The use of $V_q$ and $\chi_0$ in Eq. 3 yields the dielectric function based 
directly on the band structure of Eq. 1. The plasmon dispersion is found 
from the peak of the EELS spectrum, which is proportional to the loss 
function Im $[-1/\epsilon]$, plotted in Fig.~\ref{fig:2}. The 
theoretically predicted plasmon energy given by the maximum of the loss 
function (white trace in the figure) is compared to the 
available experimental results 
in Bi2212 (red dots).  Given that the calculation is not self-consistent, 
the agreement is seen to be reasonable. Notably, our  
calculations assume purely in-plane excitations ($q_z=0$).\cite{foot6}

We consider next the more challenging case of insulating SCOC, where 
a magnetic gap is present and it has not been clear 
how the EELS spectra\cite{WZD} are related to the plasmon excitation. 
As in the metallic case, the starting point is the dispersion of Eq. 
1, except that the hopping parameters are chosen to fit the LDA-based bands 
in SCOC: $t$=420 meV, $t'=-57$ meV, $t''$=52 meV, $t'''=26$ meV.
The correlated band structure is modeled by introducing an 
on-site Hubbard term $U$=2 eV and treating the system as a uniform 
half-filled antiferromagnet\cite{foot7} with saturated (staggered) moment 
$m_Q= 0.5$.  The band structure of Eq. 1 then splits into upper and lower 
magnetic bands with gap $\Delta_m=Um_Q$. The dielectric response 
given by Eq.~\ref{eq:13} in the metallic case is modified strongly at low 
frequencies and our analysis indicates that the dielectric function in the 
insulator can be modeled as 
\begin{equation} 
\epsilon =\epsilon_0\Bigl[1-{\omega_p^2\over \omega (\omega-i\gamma )-\omega_0^2}
\Bigr],
\label{eq:13b} 
\end{equation}
where $\omega_0\simeq U$.\cite{foot3}  Taking $\omega_0$=2 eV and the same $\gamma$ as
in Eq.~\ref{eq:13}, the EELS 
data\cite{WZD} on SCOC can be fitted to Eq.~\ref{eq:13b} with 
$\omega_p=1.3eV[1+3S_{xy}]$. 
Again, the broadening shifts the peak in Im~$[\epsilon^{-1}]$ from its expected
position $\tilde\omega_p=[\omega_p^2+\omega_0^2]^{1/2}$; the experimental peak 
can be fit to a similar $q$-dependence, but with $\omega_p^e=
2.4eV[1+0.33S_{xy}]$. Note that $\omega_p$ is
not very different in metallic Bi2212 and insulating SCOC, so that most of the 
shift in the position $\tilde\omega_p$ of the plasmon peak in the loss function 
is due to the gap $\omega_0$.

The spectral density for the dressed Green function in the insulator is 
shown in Fig.~4. The results are similar to those of Fig.~1 for Bi2212 in 
that the dressed lower magnetic band in Fig.~4 displays the waterfall 
effect and that it is renormalized in comparison to the bare band (compare 
red line with dashed blue line). The renormalization of the low energy 
spectrum by a factor of $Z\sim 0.3$ in SCOC due to the plasmon in Fig.~4 
is again comparable to the corresponding experimental value of $Z$ = 0.64.

Finally, we consider briefly the calculation of the loss function 
and the self-energy in the insulating system. With 
two bands involved in this case, the formalism is less familiar than the 
one-band metallic case, and therefore, an overview of the relvant results 
is appropriate. [The metallic case can be recovered by taking the limit 
$\Delta_m\rightarrow 0$.] The RPA expression for the charge susceptibility 
$\chi_0$ in the insulator is
\begin{equation}
\chi_0({\bf q},\omega )={1\over N}\sum_k{}' [u_-^2\bigl(g_{12}+g_{21}
         \bigr)+u_+^2\bigl(g_{11}+g_{22}\bigr)],
\label{eq:5}
\end{equation}
where the prime means summation over the magnetic Brillouin zone only,
\begin{equation}
g_{ij}^2={[1-f(\epsilon_{i,k})]f(\epsilon_{j,k+q})\over 
\epsilon_{i,k}-\epsilon_{j,k+q}+\omega+i\delta},
\label{eq:6}  
\end{equation}
$f$ is the Fermi function, 
\begin{equation}
\epsilon_{1,2k}=\epsilon_k^{(+)}\mp E_k,
\label{eq:7}  
\end{equation}
with the 1(2) going with the -(+) sign, $\epsilon_k^{(\pm )}=(\epsilon_k\pm\epsilon_{k+q})/2$, 
\begin{equation}
u_{\pm}^2={1\over 2}\Bigl[1\pm{\epsilon_k^{(-)}\epsilon_{k+q}^{(-)}+\Delta_m^2\over E_kE_{k+q}}
\Bigr],
\label{eq:8}
\end{equation}
and $E_k=[\epsilon_k^{(-)2}+\Delta_m^2]^{1/2}$.  The preceding equations 
have been applied previously to the Hubbard model\cite{ChuF,Kamp}, but we 
have extended their usage to include the long-range Coulomb interaction.

\begin{figure}  
            \resizebox{8.5cm}{!}{\includegraphics{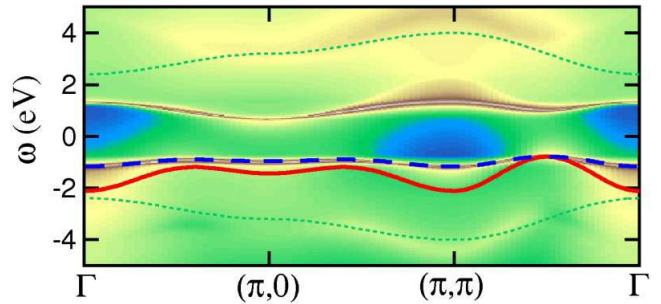}}
\vskip0.5cm
\caption{
(Color online) Dressed spectral density function in SCOC at half-filling. 
Bare dispersion (red line) is compared with dispersion renormalized with 
$Z=0.3$ (blue dashed line). Dotted lines indicate plasmon dispersion.  
Logarithmic color scale is the same as in Fig. 1.
}
\label{fig:4}
\end{figure}

\begin{figure}  
            \resizebox{8.5cm}{!}{\includegraphics{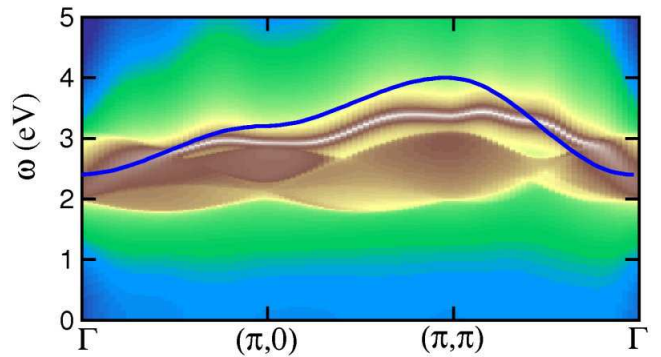}}
\vskip0.5cm
\caption{
(Color online) Loss function Im $[-1/\epsilon]$ as a function of $\omega$ 
and ${\bf q}$ in SCOC at half-filling.  Solid line gives the experimental 
plasmon dispersion ($\tilde\omega_p$) for SCOC~\protect\cite{WZD}.
Logarithmic scale as in Fig.~3. 
}
\label{fig:2b}
\end{figure}

In the presence of a gap, the electronic self energy $\Sigma$ 
becomes a tensor with indices 1(2) corresponding to 
the lower (upper)magnetic band\cite{VH,Mah}:
\begin{equation} 
\Sigma_{11}({\bf p},\omega ,\sigma )=\int_{BZ}{d^2q\over (2\pi )^2}
[v_m^2\hat\Sigma_1+u_m^2\hat\Sigma_2],
\label{eq:9b} 
\end{equation} 
\begin{equation} 
\Sigma_{12}
=\Sigma_{21}=
\int_{BZ}{d^2q\over (2\pi )^2}u_mv_m\sigma [-\hat\Sigma_1+\hat\Sigma_2],
\label{eq:9c}
\end{equation} 
where 
\begin{eqnarray} 
\hat\Sigma_i=[\Theta (\omega -\epsilon_{i,p+q})-\Theta (-\epsilon_{i,p+q})]
{V_{\bf q}^2 \chi_0({\bf q},\omega -\epsilon_{i,p+q})\over\epsilon ({\bf 
q},\omega -\epsilon_{i,p+q})/\epsilon_0}; 
\nonumber \\
u_m=\sqrt{{1\over 2}\Bigl[1+{\epsilon_{p+q}^{(-)}\over E_{p+q}}\Bigr]}; 
\ \ \ \ 
v_m=\sqrt{{1\over 2}\Bigl[1-{\epsilon_{p+q}^{(-)}\over E_{p+q}}\Bigr]},
\ \ \ \ \ \ \ \
\label{eq:12} 
\end{eqnarray}
and $\Sigma_{22}$ has the same form as $\Sigma_{11}$, but with $u_m$ and 
$v_m$ interchanged. 

The loss function in SCOC computed from Eqs. 5 and 3 is shown in Fig.~5. 
The experimentally observed plasmon dispersion (solid line) is seen to be 
in good accord with the theoretical prediction. A comparison of Figs. 3 
and 5 reveals how the gap in the underlying spectrum of the insulator 
shifts the plasmon energy in SCOC to higher values with little weight at 
energies below $\approx$ 2 eV in contrast to the metallic case of Bi2212.

The present results have significant implications for the low energy 
physics of the cuprates and provide a route for connecting this physics to 
the spectrum at higher energy scales as one encounters the venerable 
`spaghetti' of Cu and O related bands spread typically over $\sim$1-8 eV 
binding energies. We have shown that the plasmon induces band 
renormalizations at low energies in the metallic case of Bi2212 as well as 
the insulating case of SCOC, and that these renormalization effects are in 
substantial accord with the corresponding experimental values. Our results 
clearly establish the importance of the plasmon in the bosonic dressings 
of the low energy quasiparticles in the cuprates.  

There have been several recent calculations of electronically-induced 
kinks, but these are generally very weak\cite{SrVO3} or confined to a 
restricted doping range\cite{KFul}.  The plasmon by contrast is robust and 
produces strong kink-like features which extend over several eVs, due to 
the large broadening of the plasmon\cite{foot5}.  While these 
characteristic features of the electron-plasmon interaction are reminiscent of 
those reported in very recent ARPES experiments on several families of 
cuprates\cite{RonK,Ale,Non}, they fall at too high an energy scale.  
Possible candidates for the intermediate waterfall scale include the 
acoustic branch of the plasmons and 
high-energy magnetic excitations.  The extent to which low energy plasmons 
are involved in an electronic mechanism of superconductivity remains 
unclear, although such suggestions have been made in the 
literature\cite{acplas}.

In conclusion, we have shown that the low-energy plasmon, which is a 
universal feature of the cuprates, produces renormalizations and 
dispersion anomalies in the electronic structure. 
Our study indicates that the plasmon is a very significant 
player for understanding the physics of the cuprates.

{\bf Acknowledgments}: We thank B. Barbiellini for important discussions. 
This work is supported by the US DOE contract DE-AC03-76SF00098 and 
benefited from the allocation of supercomputer time at NERSC and 
Northeastern University's Advanced Scientific Computation Center (ASCC).


\begin{thebibliography}{99}
\bibitem{static}The static $\epsilon (q,0)$, calculated in RPA, screens the exchange 
potential, leading to a real, nearly $k$-independent self-energy which is presumably already 
incorporated in the LDA calculation.
\bibitem{foot6}The role of acoustic plasmons is thus underestimated. 
Our preliminary calculations suggest that this is not a serious 
limitation, but this point bears further attention.
\bibitem{foot3}The value of $\omega_0$ is estimated from the $\omega =0$ limit of 
the calculated $\epsilon (q,\omega )$.
For simplicity, the $q$-dependence of $\omega_0$ is neglected.
\bibitem{foot5}The broadening is associated with the
plasmon entering into the particle-hole continuum, and can be reproduced by adding a
small $\Sigma''$ to the bare Green's function.
\end{thebibliography}
\end{document}